\documentclass[12pt,a4paper]{article}
\usepackage{amsmath,epsfig}
\begin{document}
\begin{titlepage} 
\vspace*{0.5cm}
\begin{center}
{\Large\bf Two-loop beta functions of the Sine-Gordon model}
\end{center}
\vspace{2.5cm}
\begin{center}
{\large J\'anos Balog and \'Arp\'ad Heged\H us}
\end{center}
\bigskip
\begin{center}
Research Institute for Particle and Nuclear Physics,\\
Hungarian Academy of Sciences,\\
H-1525 Budapest 114, P.O.B. 49, Hungary\\ 
\end{center}
\vspace{3.2cm}
\begin{abstract}
We recalculate the two-loop beta functions in the two-dimensional
Sine-Gordon model in a two-parameter expansion around the
asymptotically free point. Our results agree with those of
Amit et al., J.~Phys. A13 (1980) 585.
\end{abstract}

\end{titlepage}

\newcommand{\be}{\begin{equation}}
\newcommand{\ee}{\end{equation}}
\newcommand{\bea}{\begin{eqnarray}}
\newcommand{\eea}{\end{eqnarray}}

In this paper we recalculate the two-loop beta-function 
coefficients in the two-dimensional Sine-Gordon (SG) model
in a two-parameter perturbative expansion around the
asymptotically free (AF) point. The study of the SG
model in the vicinity of this point is especially important 
since this region is used in the description of the 
Kosterlitz-Thouless (KT) phase transition in the two-dimensional
$O(2)$ nonlinear $\sigma$-model, better known as the XY 
model\footnote{
For a review of the Sine-Gordon description of the 
Kosterlitz-Thouless theory, see \cite{ZinnJ}}.
This was the motivation of the authors of \cite{Amit},
who have undertaken a systematic study of perturbation
theory in a two-parameter expansion around the AF point.
They calculated the renormalization group beta-functions
up to the two-loop coefficients. The beta-function coefficients
were also calculated in~\cite{Lovelace} by a completely 
different technique based on string theory. The results
found in~\cite{Lovelace} differ from those of~\cite{Amit}
at the two-loop level. The question of two-loop beta-function
coefficients were considered also in~\cite{Boyanovsky} for
a class of generalized Sine-Gordon models. The results,
when specialised to the case of the ordinary SG model,
agree with those of \cite{Lovelace}, but disagree with 
those of~\cite{Amit}. 
In \cite{LeClair} the short distance expansion of some Sine-Gordon  
correlation functions were calculated using conformal perturbation 
theory. This allowed the extraction of the one- and two-loop 
beta-function coefficients around the AF point. The resulting 
two-loop beta-functions differ from all the previous results.

In view of the role the Sine-Gordon model is playing in the 
description of the KT phase transition it is very important to know 
the correct two-loop beta-function coefficients.
The purpose of the present paper is to show that, in fact,
the two-loop results of Amit et al. are the correct ones. We show this
first by comparing the SG beta-function 
to known results in the chiral Gross-Neveu model \cite{Destri},
which is known to be equivalent to the SG model at its 
AF point. We also check the beta-function coefficients by
considering the renormalization of $2n$-point functions
of exponentials of the SG field.

Following \cite{Amit} we consider the Euklidean Lagrangian
\begin{equation}
{\cal L}=\frac12\partial_\mu\phi\partial_\mu\phi
+\frac{m_0^2}{2}\phi^2+\frac{\alpha_0}{\beta_0^2a^2}\left[
1-\cos(\beta_0\phi)\right],
\label{L}
\end{equation}
where $m_0$ is an IR regulator mass and $a$ is the UV
cutoff (of dimension length). UV regularized correlation
functions are calculated by using
\begin{equation}
G_0(x)=\frac{1}{2\pi}\,K_0\left(m_0\sqrt{x^2+a^2}\right)
\label{G0}
\end{equation}
where $K_0$ is the modified Bessel function, as the 
$\phi$ propagator. Our strategy is slightly different from
\cite{Amit}, who really considered the renormalization of
the massive SG model (\ref{L}) of mass $m_0$. We treat
$m_0$ as an IR regulator mass and consider IR stable
physical quantities for which we can take the limit 
${m_0\to0}$ already at the UV regularized level
(before UV renormalization).

The model (\ref{L}) is renormalizable in a two-parameter
perturbative expansion around the point corresponding to
the couplings $\alpha_0=0$, $\beta_0^2=8\pi$. Writing
\begin{equation}
\beta_0^2=8\pi(1+\delta_0)
\label{delta0}
\end{equation}
the two bare expansion parameters are $\alpha_0$ and
$\delta_0$ and physical quantities can be made UV finite
by the renormalizations
\begin{eqnarray}
\alpha_0&=&Z_\alpha\,\alpha;\,Z_\alpha=1+g_1\delta\ell+
\alpha^2(\overline{g}_2\ell^2+g_2\ell)+
\delta^2(\overline{g}_3\ell^2+g_3\ell)+\dots,
\label{Zalpha}\\
1+\delta_0&=&Z_\phi^{-1}(1+\delta);\qquad
Z_\phi=1+f_1\alpha^2\ell+
\alpha^2\delta(\overline{f}_2\ell^2+f_2\ell)
+\dots,\label{Zphi}
\eea
where $\alpha$ and $\delta$ are the renormalized couplings
and $\ell=\ln\mu a$ with $\mu$ an arbitrary renormalization
point. The dots stand for terms higher order in perturbation
theory and the numerical coefficients $g_1,f_1$ etc. can be
calculated by renormalizing correlation functions. 
The results of Amit et al. are
\begin{equation}
f_1={\textstyle\frac{1}{32}},\quad
g_1=-2,\quad
f_2=-{\textstyle\frac{3}{32}},\quad
g_2=-{\textstyle\frac{5}{64}},\quad
g_3=0,
\label{ResAmit}
\end{equation}
those of \cite{Lovelace} and \cite{Boyanovsky} are
\begin{equation}
f_1={\textstyle\frac{1}{32}},\quad
g_1=-2,\quad
f_2=-{\textstyle\frac{1}{32}},\quad
g_2=-{\textstyle\frac{1}{32}},\quad
g_3=0
\label{ResLovelace}
\end{equation}
and finally \cite{LeClair} found
\begin{equation}
f_1={\textstyle\frac{1}{32}},\quad
g_1=-2,\quad
f_2=-{\textstyle\frac{1}{32}},\quad
g_2=-{\textstyle\frac{1}{16}},\quad
g_3=0.
\label{ResKonik}
\end{equation}
We see that the one-loop coefficients are the same but not all
two-loop coefficients agree. The subject of this paper is to 
recalculate these numbers.

The renormalization group (RG) beta-functions can be calculated
by solving the equations
\begin{equation}
{\cal D}\alpha={\cal D}\delta=0,
\label{RG}
\end{equation}
where, as usual, the RG operator is defined by
\begin{equation}
{\cal D}=-a\frac{\partial}{\partial a}+
\beta_\alpha(\alpha_0,\delta_0)\frac{\partial}
{\partial\alpha_0}
+\beta_\delta(\alpha_0,\delta_0)\frac{\partial}
{\partial\delta_0}.
\label{D}
\end{equation}
One finds
\bea
\beta_\alpha&=&-g_1\alpha_0\delta_0-g_2\alpha_0^3-
g_3\alpha_0\delta_0^2+\dots,\label{betaalpha}\\
\beta_\delta&=&f_1\alpha_0^2+(f_1+f_2)\alpha_0^2\delta_0
+\dots\label{betadelta}
\eea

It is well-known that, in the case of several couplings,
the higher beta-function coefficients are not all scheme
independent. Indeed, considering the perturbative
redefinitions
\begin{equation}
\tilde{\alpha}_0=\alpha_0+c_1\alpha_0\delta_0+\dots\qquad
\tilde{\delta}_0=\delta_0+c_2\alpha_0^2+\dots
\label{redef}
\end{equation}
one finds that in addition to the one loop coefficients
$f_1$ and $g_1$ only the following two two-loop coefficient 
combinations are invariant.
\begin{equation}
g_3,\qquad J=2g_2-f_2.
\label{invariants}
\end{equation}

The RG analysis with two couplings can be made similar to the 
case of a single coupling by changing the variables from
$\alpha_0$ and $\delta_0$ to the pair $Q$ and $\delta_0$,
where $Q$ is a RG invariant (solution of the ${{\cal D}Q=0}$
equation), given in perturbation theory by
\begin{equation}
Q=f_1\alpha_0^2+g_1\delta_0^2+2g_2\alpha_0^2\delta_0+F_2
\delta_0^3+\dots,
\label{Q}
\end{equation}
where $F_2=\frac23g_3-\frac23g_1+\frac{2g_1}{3f_1}J$.
Now $Q$, being a RG invariant, can almost be treated as
if it were a numerical constant and $\delta_0$ as the
\lq true' coupling. The beta-function in these variables is
\begin{equation}
\beta(\delta_0,Q)=Q+2\delta_0^2+AQ\delta_0+B\delta_0^3+\dots,
\label{betaQ}
\end{equation}
where $A=1-J/f_1$ and $B=2(A-g_3)/3$.

It is well-known that the SG model can also be formulated
in terms of two fermion fields, interacting with a chirally
symmetric current-current interaction \cite{ZinnJ}. A special
case of the two-fermion model corresponds to the 
$SU(2)$-symmetric chiral Gross-Neveu model. This correspondence
is evident in the bootstrap aproach, since the SG S-matrix
in the limit ${\beta_0\to\sqrt{8\pi}}$ becomes the $SU(2)$
chiral Gross-Neveu S-matrix. This asymptotically free model
has to correspond to one of the possible RG trajectories
in the two-parameter SG language. It is easy to see that
it has to be the $Q=0$ trajectory, since this is the only
trajectory going through the origin (${\delta_0=\alpha_0=0}$)
of the parameter space. More precisely, the chiral Gross-Neveu
model must correspond to the negative half of the $Q=0$
trajectory, which is an UV asymptotically free trajectory.
Making the identification 
\begin{equation}
\delta_0=-\frac{1}{\pi}g^2,
\label{g}
\end{equation}
where $g$ is the coupling of the $SU(2)$ Gross-Neveu model,
the Gross-Neveu beta-function becomes
\begin{equation}
\beta(g)=-\frac{1}{\pi}g^3+\frac{B}{2\pi^2}g^5+\dots
\label{betaGN}
\end{equation}
Using the results of Amit et al., (\ref{ResAmit}), $B=2$,
using the results of \cite{Lovelace} and 
\cite{Boyanovsky}, (\ref{ResLovelace}), $B=4/3$ and finally 
$B=8/3$ if we trust \cite{LeClair}, (\ref{ResKonik}). 
Comparing (\ref{betaGN})
to the results of the beta-function calculations performed
directly in the fermion language \cite{Destri} we see that
the correct Gross-Neveu beta-function is reproduced if
$B=2$. Thus the two-loop results of Amit et al. are correct
after all! This was the observation\footnote{We thank 
P. Forg\'acs who made this observation first and called
our attention to it.} that served as our motivation for 
the present study. The correctness of the two-loop Gross-Neveu
beta-function coefficient has been checked by studying the
system in the presence of an external field \cite{Forgacs}.
Using this method the value of this coefficient can be read 
off from the bootstrap S-matrix and the results are in agreement
with \cite{Destri}.

We now turn to the explicit calculation of the 
renormalization parameters (\ref{Zalpha},\ref{Zphi}).
The first quantity we consider is the two-point function
of the $U(1)$ current\break
 ${J_\mu=i\frac{\beta_0}{2\pi}
\epsilon_{\mu\nu}\partial_\nu\phi}$,
\begin{equation}
\langle J_\mu(x)J_\nu(y)\rangle=
\int\frac{d^2p}{(2\pi)^2}\left(\frac{p_\mu p_\nu}{p^2}-
\delta_{\mu\nu}\right)e^{ip(x-y)}I(p).
\label{JJ}
\end{equation}
The advantage of considering this physical quantity is that 
it is IR stable. Putting $m_0=0$ we find
\begin{equation}
I(p)=\frac{2}{\pi}\left\{1+\delta_0+\frac{\alpha_0^2}{32}
\left(\ln pa+K+\frac12\right)+\frac{\alpha_0^2\delta_0}{16}
(\ln pa+K)^2+\dots\right\},
\label{I}
\end{equation}
where $K=-\Gamma^\prime(1)-1-\ln2$. Since the current is 
conserved there is no operator renormalization required here
and (\ref{I}) must become finite after the substitutions 
(\ref{Zalpha},\ref{Zphi}). From this requirement we get
\begin{equation}
f_1={\textstyle\frac{1}{32}},\quad
g_1=-2,\quad
f_2=-{\textstyle\frac{3}{32}}.
\label{ResI}
\end{equation}

To determine the remaining two-loop coefficients $g_2$ and
$g_3$ we have to calculate $Z_\alpha$, the renormalization constant 
corresponding to $\alpha_0$. For this purpose we need a
quantity with a perturbative series starting at ${\cal O}(\alpha_0)$.
We have chosen the $2n$-point correlation function 
\begin{equation}
X=\langle
{\cal A}(x_1)\dots
{\cal A}(x_{2n})\rangle,
\label{X}
\end{equation}
where
\begin{equation}
{\cal A}(x)=\left(\frac{1}{a}\right)^{\frac{1}{2n^2}}e^{\frac{i\beta_0}
{2n}\phi(x)}.
\label{A}
\end{equation}
Although, in contrast to the Noether current, the operator 
\eqref{A} needs to be renormalized, for large enough
$n$ the dimension of \eqref{A} is so small that there is no operator
mixing and the operator renormalization constant can simply be determined
from the correlation function
\begin{equation}
Y=\langle
{\cal A}(x_1)\dots
{\cal A}(x_n)
{\cal A}^*(y_1)\dots
{\cal A}^*(y_n)\rangle.
\label{Y}
\end{equation}
A second order calculation gives
\begin{equation}
\begin{split}
Y=&M^{\left(\frac{1}{n^2}\right)}\Big\{1
+\frac{\delta_0}{n^2}L
+\frac{\delta_0^2}{2n^4}L^2
+\frac{\alpha_0^2}{64n^3}L^2\\
&+L\left(\frac{\alpha_0^2}{64n^2}
-\frac{\alpha_0^2}{128}\left[
W\left(\frac{1}{n}\right)+
W\left(-\frac{1}{n}\right)\right]\right)+\dots\Big\},
\label{Yres}
\end{split}
\end{equation}
where
\begin{equation}
M=\frac{
\prod_{i<j}\vert x_i-x_j\vert\, \prod_{k<l}\vert y_k-y_l\vert}
{\prod_{i,k}\vert x_i-y_k\vert},
\label{M}
\end{equation}
$L=\ln Ma^n$ and the dots stand for finite ${\cal O}(\alpha_0^2)$ terms
as well as higher order terms. $W(\mu)$ is defined by
\begin{equation}
W(\mu)=-1
+\int_0^1dz\,z^\mu F(\mu,\mu;1;z)
+\int_0^1\frac{dz}{z^2}\left[F(\mu,\mu;1;z)-1-\mu^2z\right],
\label{W}
\end{equation}
where $F(\alpha,\beta;\gamma;z)$ is the standard hypergeometric function.
\eqref{Yres} can be made finite by the renormalization $Y_R=Z_{2n}Y$,
where
\begin{equation}
Z_{2n}=1-\frac{1}{n}\ell\delta+\frac{1}{2n^2}\ell^2\delta^2+
\frac{1}{64n}\ell^2\alpha^2+k_1\ell\alpha^2+\dots,
\label{zeta}
\end{equation}
with
\begin{equation}
k_1=-\frac{1}{64n}+\frac{n}{128}\left[W\left(\frac{1}{n}\right)
+W\left(-\frac{1}{n}\right)\right].
\label{k1}
\end{equation}

For the $2n$-point function $X$ a second order calculation gives
\begin{equation}
\begin{split}
X=&\frac{\alpha_0}{16\pi}N^{\left(\frac{1}{n^2}\right)}\,F\,\Bigg\{
1+\delta_0\Psi+\frac12\delta_0^2\Psi^2\\
&+\frac{n\alpha_0^2}{128n+64}
\Psi\left[\Psi+4+\frac{1}{n}-nW\left(\frac{1}{n}\right)\right]+\dots
\Bigg\},
\end{split}
\label{Xres}
\end{equation}
where
\begin{equation}
N=\prod_{i<j}\vert x_i-x_j\vert,
\qquad\qquad
F=\int d^2z\frac{1}{\prod_i\vert z-x_i\vert^{\frac{2}{n}}}
\label{NF}
\end{equation}
and
\begin{equation}
\Psi=-1+\frac{1}{n^2}\ln \left(Na^{-n(2n-1)}\right)-\frac{2}{nF}
\sum_j\int
d^2z\frac{\ln\left\vert\frac{z-x_j}{a}\right\vert}{\prod_{i}\vert
z-x_i\vert^{\frac{2}{n}}}.
\label{}
\end{equation}
In \eqref{Xres} the dots represent finite terms of ${\cal O}(\alpha_0^2)$
and ${\cal O}(\delta_0^2)$ as well as higher terms.
Renormalizing $X$ by requiring $X_R=Z_{2n}X$ to be finite after
coupling constant renormalization gives 
\begin{equation}
g_3=0\qquad{\rm and}\qquad
g_2=-\frac{1}{16}+\frac{n}{128}\left[
W\left(\frac{1}{n}\right)-W\left(-\frac{1}{n}\right)\right].
\label{g32}
\end{equation}
At first sight $g_2$ seems to be $n$-dependent which would mean that
the $2n$-point function \eqref{X} cannot really be made finite with wave
function plus coupling constant renormalization. Luckily, however,
one can see that using the identity
\begin{equation}
W(\mu)-W(-\mu)=-2\mu\qquad\qquad(\vert\mu\vert<1)
\label{identity}
\end{equation}
satisfied by the hypergeometric function, $g_2$ is equal to the
$n$-independent constant $-5/64$. Moreover, \eqref{g32} together with
\eqref{ResI} reproduce \eqref{ResAmit}, the results of \cite{Amit}.
The nontrivial cancellation of the $n$-dependence makes us 
more confident that these are the correct two-loop coefficients.
\bigskip

\noindent{\it Acknowledgements}

\noindent
We would like to thank P. Forg\'acs for calling our attention to the fact
that the results of \cite{Amit} reproduce the correct two-loop coefficients
for the chiral Gross-Neveu model. We thank R. Konik for a correspondence.


\end{document}